\documentclass[12pt]{article}
\input epsf.sty

\begin{document} 
 
\def\CA{{\cal A}} 
\def\CB{{\cal B}} 
\def\CC{{\cal C}} 
\def\CD{{\cal D}} 
\def\CE{{\cal E}} 
\def\CF{{\cal F}} 
\def\CG{{\cal G}} 
\def\CH{{\cal H}} 
\def\CI{{\cal I}} 
\def\CJ{{\cal J}} 
\def\CK{{\cal K}} 
\def\CL{{\cal L}} 
\def\CM{{\cal M}} 
\def\CN{{\cal N}} 
\def\CO{{\cal O}} 
\def\CP{{\cal P}} 
\def\CQ{{\cal Q}} 
\def\CR{{\cal R}} 
\def\CS{{\cal S}} 
\def\CT{{\cal T}} 
\def\CU{{\cal U}} 
\def\CV{{\cal V}} 
\def\CW{{\cal W}} 
\def\CX{{\cal X}} 
\def\CY{{\cal Y}} 
\def\CZ{{\cal Z}} 
 
\newcommand{\btzm}[0]{BTZ$_{\rm M}$} 
\newcommand{\todo}[1]{{\em \small {#1}}\marginpar{$\Longleftarrow$}} 
\newcommand{\labell}[1]{\label{#1}\qquad_{#1}} 
\newcommand{\ads}[1]{{\rm AdS}_{#1}} 
\newcommand{\SL}[0]{{\rm SL}(2,\IR)} 
\newcommand{\cosm}[0]{R} 
\newcommand{\tL}[0]{\bar{L}} 
\newcommand{\hdim}[0]{\bar{h}} 
\newcommand{\bw}[0]{\bar{w}} 
\newcommand{\bz}[0]{\bar{z}} 
\newcommand{\be}{\begin{equation}} 
\newcommand{\ee}{\end{equation}} 
\newcommand{\lp}{\lambda_+} 
\newcommand{\bx}{ {\bf x}} 
\newcommand{\bk}{{\bf k}} 
\newcommand{\bb}{{\bf b}} 
\newcommand{\BB}{{\bf B}} 
\newcommand{\tp}{\tilde{\phi}} 
\hyphenation{Min-kow-ski} 
 
\def\ie{{\it i.e.}} 
\def\eg{{\it e.g.}} 
\def\cf{{\it c.f.}} 
\def\etal{{\it et.al.}} 
\def\etc{{\it etc.}} 
 
\def\apr{\alpha'} 
\def\str{{str}} 
\def\lstr{\ell_\str} 
\def\gstr{g_\str} 
\def\Mstr{M_\str} 
\def\lpl{\ell_{pl}} 
\def\Mpl{M_{pl}} 
\def\varep{\varepsilon} 
\def\del{\nabla} 
\def\grad{\nabla} 
\def\tr{\hbox{tr}} 
\def\perp{\bot} 
\def\half{\frac{1}{2}} 
\def\p{\partial} 
 
\renewcommand{\thepage}{\arabic{page}} 
\setcounter{page}{1}

\rightline{HUTP-99/A032, UCSBTH-99-1} 
\rightline{hep-th/9906226} 
\vskip 1cm 
\centerline{\Large \bf Holographic Particle Detection} 
\vskip 1cm 
 
\renewcommand{\thefootnote}{\fnsymbol{footnote}} 
\centerline{{\bf Vijay 
Balasubramanian${}^{1}$\footnote{vijayb@pauli.harvard.edu} and 
Simon F. Ross${}^{2}$\footnote{sross@cosmic.physics.ucsb.edu},}}  
\vskip .5cm 
\centerline{${}^1$\it Jefferson Laboratory of Physics, Harvard
University,}  
\centerline{\it Cambridge, MA 02138, USA} 
\vskip .5cm 
\centerline{${}^2$ \it Department of Physics,} 
\centerline{\it University of California,} 
\centerline{\it Santa Barbara, CA 93106, USA} 
 
\setcounter{footnote}{0} 
\renewcommand{\thefootnote}{\arabic{footnote}} 
 
\begin{abstract} 
In anti-de Sitter (AdS) space, classical supergravity solutions are
represented ``holographically" by conformal field theory (CFT) states
in which operators have expectation values. These 1-point functions
are directly related to the asymptotic behaviour of bulk fields.  In
some cases, distinct supergravity solutions have identical asymptotic
behaviour; so dual expectation values are insufficient to distinguish
them. We argue that non-local objects in the gauge theory can resolve
the ambiguity, and explicitly show that collections of point particles
in $\ads{3}$ can be detected by studying kinks in dual CFT Green
functions.  Three dimensional black holes can be formed by collision
of such particles.  We show how black hole formation can be detected
in the holographic dual, and calculate CFT quantities that are
sensitive to the distribution of matter {\it inside} the event
horizon.
\end{abstract} 
 
\section{Introduction} 
  
According to the ``holographic principle''~\cite{holog}, quantum
gravity in spacetimes with some prescribed asymptotic behaviour can be
described by a theory defined on the boundary at infinity. Such a
principle implies an enormous reduction in degrees of freedom relative
to conventional local quantum field theory in the bulk spacetime.
Nevertheless, everyday experience shows that local objects exist in
the semiclassical limit.  So, in order to accept the holographic
principle, it is necessary to understand how such objects are encoded
and detected in the holographic dual.
 
A simple example of holography is the proposed duality between gravity 
in anti-de Sitter (AdS) spaces and local conformal field theories 
(CFTs) defined on the AdS boundary~\cite{juanads}.  The semiclassical 
limit in AdS is related to a large central charge limit of the 
CFT.  In this context, classical supergravity probes are described by 
dual CFT states in which operators have expectation values derived 
directly from the asymptotic behaviour of fields coupling to the 
probe~\cite{bklt}.  This was used to show a ``scale-radius duality'' 
for a variety of bulk sources, and for wavepackets of supergravity 
fields -- the radial position of a bulk probe is encoded in the scale 
size of the dual expectation values.  Dynamical sources for 
supergravity fields were studied in~\cite{dkk}, where the radial 
position of a source particle following a bulk geodesic was reflected 
in the size and shape of an expectation-value bubble in the CFT. 
Other interesting work along these lines was performed 
in~\cite{bdhm}-\cite{joeetal}. 
 
The scale-radius duality provides a simple example of the holographic 
encoding of information about the location and motion of objects in 
the additional spacetime dimension. However, as we will discuss in 
Sec.~\ref{1pt}, the simple scale-radius relationship is a consequence 
of an isometry in pure AdS under rescalings of all coordinates. 
For spacetimes such as black holes which break the isometries, the 
relationship between bulk position and boundary observables will be 
more complicated.  We will illustrate this point by discussing a 
dilaton source falling into a BTZ black hole, following Danielsson 
et.al.~\cite{dkk}.  The same phenomenon is apparent in the collision 
of two particles to form a black hole in~\cite{joeetal}; after the 
particles collide, their radial position is fixed, but the scales in 
the boundary expectation values continue to evolve. 
 
If a simple scale-radius duality fails, do CFT expectation values
still tell us the location of bulk sources? In Sec.~\ref{1pt}, we will
review the surprising power of the expectation values, and discuss
what one can learn about the bulk in general. If we write AdS in the
global coordinate system, all the normalizable mode spherical
harmonics fall off at the same rate at the spacetime boundary. As a
result, expectation values of the dual operators contain all the
multipole moments of any bulk solution which is constructed by
superposition of these modes. Therefore, such supergravity
configurations will be completely determined by the CFT one-point
functions.
 
However, there are distinct supergravity configurations which are
identical outside of some region. Examples include spacetimes with
spherically symmetric matter distributions, and collections of point
particles in $2+1$ dimensions.\footnote{The matter sources in these
solutions must have sharp cutoffs -- otherwise they would be detected
in dual 1-point functions as above.  Thus, they are not made from the
supergravity fields, all of which have tails at infinity. One might
perhaps think of them as made up of massive strings.}  It is important
to describe these examples from the CFT perspective, because they
include simple processes of particular interest, such as spherically
symmetric collapse to form a black hole.  In such cases, CFT
expectation values will not resolve distinct bulk configurations.  We
will argue that there are non-local quantities in the gauge theory
that can provide the necessary data.
 
The core of this paper is the use of the Feynman Green function to 
extract information about bulk solutions from the CFT. We consider 
asymptotically $\ads{3}$ spacetimes containing point particles, as 
this is a particularly tractable example. In Sec.~\ref{single}, we 
derive the effect of a stationary particle on the Green function of an 
operator dual to a scalar field. We compute the Green function by a 
WKB approximation, relating it to the length of geodesics, and find 
that the particle's presence is signalled by a kink in the propagator. 
In Sec.~\ref{moving}, we show how the motion of particles in $\ads{3}$ 
is reflected in these kinks. 
 
In $\ads{3}$, the head-on collision of particles produces a BTZ black
hole~\cite{BTZ} if the combined energy exceeds a certain threshold.
We show in Sec.~\ref{two} that in the holographic description of this
process there are initially two kinks in the CFT Green function,
indicating the separate positions of the two particles. The
discontinuities approach each other and merge into a single one
associated with the final black hole.  Remarkably, the kinks are
sensitive to the positions of the particles {\em inside} the event
horizon, until their eventual collision to form a singularity.  This
supports the assertion that the holographic dual gives a unitary
description of processes localized inside a black hole horizon, which
should have important implications for information loss. Flat space
may be obtained from AdS by taking the length scale ($\ell$) of the
spacetime to infinity.  The measurement precision required for
detecting the separation of two particles increases with $\ell$.  In
the flat space limit, we find that infinite precision measurements are
required in the holographic dual to three dimensional gravity to
detect the difference between one and two bulk particles.
 
We conclude in Sec.~\ref{conc} by drawing some general lessons from 
our work and suggesting directions for the future. 
 
\section{Classical probes and dual expectation values} 
\label{1pt} 
 
In the Lorentzian version of the AdS/CFT correspondence, we must
specify both boundary conditions at infinity and initial conditions
for the normalizable modes in spacetime.  Changing the boundary
conditions at infinity corresponds to turning on sources in the dual
CFT, while the choice of normalizable modes corresponds to the
state~\cite{bkl}. In~\cite{bklt}, it was shown that both of these
choices will affect the one-point functions of the operators dual to
supergravity fields, and the effect of a number of bulk sources on the
expectation values was explicitly calculated. Here, we wish to make
some general comments, drawing on previously studied examples for
support.  The expectation value of the CFT operator is essentially
given by the asymptotic value of the normalizable modes~\cite{bkl}. In
Poincar\'e coordinates,
\begin{equation}  
\langle {\cal O}(\bx) \rangle = \Delta \,\tilde{\phi}_n(\bx) + c \int 
d^d \bx' {\phi_0({\bf x}') \over |\bx - \bx'|^{2\Delta}}. 
\label{onept} 
\end{equation} 
The first term is the effect of a normalizable mode $\phi_n(z,\bx)$, 
behaving as $z^{\Delta}\, \tilde{\phi}_n(\bx)$ as $z \to 0$, and the 
second term is the effect of a change in the boundary conditions by 
$\phi_0(\bx')$. 
 
The normalizable modes of scalar fields of mass $ m$ in global 
AdS$_{d+1}$ with scale $\ell$ behave as~\cite{bkl} 
\begin{equation}  
\Phi = e^{-i \omega t} \, Y_{l,\{{\bf m}\}}(\Omega) \, (\cos 
\rho)^{\Delta} \, (\sin \rho)^l \, {}_2 F_1(a,b,c;\cos^2 \rho), 
\label{globalnorm} 
\end{equation} 
where the boundary is at $\rho = \pi/2$, $\Delta = (d
+\sqrt{d^2+4m^2\ell^2})/2$, and $Y_{l,\{ {\bf m}\}}(\Omega)$ denotes
spherical harmonics on $S^{d-1}$.  The falloff of the modes as $\rho
\to \pi/2$ is {\it independent} of the angular quantum numbers. Thus,
the dual expectation values $\langle {\cal O}(t,\Omega) \rangle$ will
determine all the spherical harmonics, and therefore all the multipole
moments, of the bulk solution.\footnote{In $\ads{2}$, the bulk
spacetime has two boundaries; however, specifying the boundary
conditions and expectation values on one boundary is sufficient to
determine the bulk solutions, and hence the values on the other
boundary.  Therefore the $\ads{2}$ theory has only one set of
independent observables, corresponding to the expected dual quantum
mechanics.}
 
This ensures that the expectation values will contain all information 
about any solution constructed by superposition of the basic modes 
(\ref{globalnorm}). At infinity, the fields fall off exponentially in 
proper distance, but the exponential is the same for all modes. 
Although the bulk fields are themselves minuscule at the boundary, CFT 
expectation values are given by the finite prefactor to the decaying 
exponential. 
  
\subsection{Scale-radius duality: the power of symmetry} 
 
As seen in the examples discussed in~\cite{bklt, bdhm}, and in several 
subsequent papers, the radial position of supergravity sources in pure 
AdS is encoded in the dual one-point function in a particularly simple 
way; radial translation of the source corresponds to a rescaling of 
the corresponding expectation value. This is called scale-radius 
duality, and is an example of a general relationship between radial 
positions and boundary scales~\cite{susswitt, amanda}. 
 
For sources at fixed Poincar\'e coordinate radial positions, this 
relation follows because the AdS metric is unchanged by the rescaling 
\begin{equation}  
\bx \to \lambda \bx, ~~~~ z \to \lambda z. 
\label{resc} 
\end{equation} 
Given the solution describing a source at some radial position, we can
describe a source at a different radial position by the redefinition
(\ref{resc}).  The effect on the asymptotic fields, and hence the
one-point function, is simply a rescaling of the
coordinates~\cite{bklt}.
 
In global AdS, symmetry dictates that a source at the origin is
represented by a CFT expectation value which is constant over the
boundary sphere. There is an isometry mapping this source to one
following any other geodesic in the spacetime. This isometry acts as a
conformal transformation in the dual CFT, a fact exploited
in~\cite{dkk,garysunny} to obtain CFT expectation values dual to
moving sources in the bulk.
 
In cases with less symmetry, the story is more complicated.  A good
example is a point source for the dilaton falling into a BTZ black
hole, which was studied by Danielsson et.al.~\cite{dkk} using the
method of images and the result for sources in $\ads{3}$.  In the
lightlike limit, such a particle entering the boundary at $t=0$,
produces the asymptotic dilaton field
\begin{equation}  
\Phi = \sum_{n=-\infty}^{\infty} {\delta(\phi_+ +2\pi n) \over 
\sinh^{\nu+1} [{(r_+-r_-) \over 2}(\phi_- - 2\pi n)]} + {\delta(\phi_- - 
2\pi n) \over \sinh^{\nu+1}[{(r_+-r_-) \over 2}(\phi_+ + 2\pi n)]}, 
\label{eskobtz} 
\end{equation} 
where $\phi_{\pm}= t\pm\phi$ and $r_{\pm}$ are the
horizons.\footnote{Our source particle has a slightly different
normalization from the one in~\cite{dkk}.}  Therefore, the
corresponding expectation value is concentrated on the light cones $t
= \pm \phi$, as for a single particle in AdS$_5$~\cite{garysunny}. The
sum over $n$ implements the periodic identification of $\phi$ in the
BTZ black hole.  Simple scale-radius duality fails here, in the sense
that the two light cones cross repeatedly; so the scale size goes to
zero, e.g. at $t=\pi$, while the particle never returns to the
boundary. In fact, the information concerning the location of the
particle in the bulk seems to be contained in the amplitude of $\Phi$,
which steadily declines in time.
 
A related example is a pair of particles colliding to form a black 
hole -- the boundary expectation value associated with each particle 
will continue to evolve independently after the collision, reaching zero 
scale size at a later time~\cite{joeetal}.  In this case also, the 
isometries used above are not available, being broken by the metric in 
the region to the future of both particles.  Consequently, the simple 
connection between scale size and radial position fails to apply 
directly. 
 
\subsection{Resolving bulk objects with propagators} 
 
There are also examples 
where the CFT expectation values implied by (\ref{onept}) do not fully 
characterize a semiclassical bulk state.  For example, a spherical 
shell of matter placed at any position within AdS will have the same 
asymptotic fields and dual 1-point functions.  (Requiring the matter 
distribution to have compact support implies that these objects are 
not built from the supergravity modes (\ref{globalnorm}), all of which 
have tails at infinity.) Similarly, the only supergravity data 
available at infinity about a collection of point particles in 
$\ads{3}$ is the total mass.  So, the dual CFT stress tensor has an 
expectation value~\cite{nsetal,mart,adsstress}, but this is 
insufficient to count the particles or locate them in spacetime. 
 
Nevertheless, the CFT should somehow encode the difference between
these bulk configurations.  We wish to identify some CFT quantities
that probe the interior of the spacetime and characterize the bulk
solutions conveniently.  These quantities should be non-local from the
CFT perspective, because the scale-radius duality teaches us that
locations further inside the spacetime correspond to larger scales in
the gauge theory.  Wilson loop expectation values are such quantities
-- they depend on the action of strings that start on the boundary and
penetrate the spacetime~\cite{wilson}.  We can also use scattering
amplitudes for the operators dual to supergravity fields.  As we will
see, in the WKB approximation the CFT propagator is dominated by bulk
geodesics.  Since these penetrate the spacetime, they can be used to
distinguish bulk configurations with identical asymptotic fields.
Recently this approach has been applied to spherical shells of matter
in $\ads{5}$~\cite{dkk2}, which will be analyzed further in a
forthcoming paper involving one of us~\cite{stevesimon}.  In the
remainder of this paper we will work out the details of this procedure
for resolving point particles in three dimensions.

\section{Stationary particles in $\ads{3}$} 
\label{single} 
 
Global $\ads{3}$ is described by the metric 
\begin{equation} 
ds^2  = d\chi^2 + \sinh^2\chi \, d\phi^2 - \cosh^2\chi 
\, dt^2 = \left({2 \over 1- r^2}\right)^2 (dr^2 + r^2 d\phi^2) - 
\left( {1 + r^2 \over 1 - r^2} \right)^2 dt^2. 
\label{adsmet} 
\end{equation} 
We have set the AdS length scale $\ell$ to one, $\phi$ has a period
$2\pi$, time runs between $\pm\infty$ and $0 \leq r = \tanh(\chi/2)
\leq 1$ is the radial coordinate.  Fixed $t$ surfaces have the
Poincar\'e disc geometry, and the dual CFT is defined on the cylinder
at the $r = 1$ boundary.  The mass of $\ads{3}$ may be computed as
in~\cite{adsstress} to be $M = - 1/8 G$ and is equal to the ground
state energy of the dual conformal field theory~\cite{strom}.  We will
introduce point particles into this spacetime and will show that kinks
in the propagator for CFT operators in the dual state locate the
particle in the spacetime.  It is helpful to begin by recasting the
computation of CFT correlators from $\ads{3}$ in a language that is
convenient for generalization to point particle states.

\subsection{$\ads{3}$ geodesics and dual propagators} 
\label{geodcorr} 
 
A generic scalar field of mass $m$ in asymptotically $\ads{3}$ spaces 
is dual to an operator $\CO$ of conformal dimension $\Delta = 1 + 
\sqrt{1 + m^2}$.  We are going to show how the properties of 
point particles in $\ads{3}$ can be read off from the Green 
functions of such operators.  It is helpful to begin by computing the 
propagator in the conformally invariant vacuum.  We can normalize the 
operator $\CO$ so that on the Euclidean plane ($ds^2 = dx_1^2 + dx_2^2 
= dr^2 + r^2 d\phi^2$) its 2-point function is constrained by 
conformal invariance to be 
\begin{equation} 
\langle \CO({\bf x}) \CO({\bf x}') \rangle = 
{1 \over | {\bf x} - {\bf x}'|^{2\Delta}} = 
{1 \over (r^2 + r^{\prime 2} - 2 r r' \cos\phi)^\Delta}, 
\end{equation} 
where $r$ and $r'$ are the distances from the origin and $\phi$ is the 
angle between ${\bf x}$ and ${\bf x}'$.    Transform to the 
Euclidean cylinder ($ds^2 = dt^2 + d\phi^2$) by setting $r = e^t$ and 
Weyl rescaling the metric by $e^{-2t}$.  Let $-\pi \leq z \leq \pi$ 
parametrize any closed curve on this cylinder with the condition 
\begin{equation} 
\bb(z) \equiv(t(z),\phi(z))= (t(-z),-\phi(-z)) 
\label{cond} 
\end{equation} 
imposed for later convenience.  Then  
\begin{equation} 
T(z) \equiv \ln \langle \, \CO(\bb(z))\, \CO(\bb(-z)) \, \rangle 
= -2\Delta \, \ln(\, 2 \, \sin\phi(z)\,). 
\label{eqt1} 
\end{equation} 
Wick rotating to Lorentzian signature ($ds^2 = -dt^2 + d\phi^2$)
leaves (\ref{eqt1}) unchanged, but in Lorentzian signature it should
be understood as the logarithm of the Feynman Green function for
$\CO$. We want to reproduce this from the AdS/CFT correspondence in a
manner convenient for generalization to point particles in $\ads{3}$.
 
The CFT dual to $\ads{3}$ is defined on a cylinder with diverging Weyl 
factor.  So the Green function for dual operators at any finite 
coordinate separation will vanish.  It is convenient to regulate this 
behaviour by cutting off the spacetime at a boundary defined by 
\begin{equation} 
r_{{\rm m}}(t,\phi) = 1 - \epsilon(t,\phi),~~~~~ 
\epsilon(t,\phi) = \epsilon(t,-\phi), 
\label{regulate} 
\end{equation} 
where $\epsilon$ is some smooth function of the boundary coordinates.
The symmetry of $\epsilon$ under $\phi \rightarrow -\phi$ is merely
chosen to reduce verbiage in the remainder of this paper, and ensures
that boundary curves $\bb(z)$ satisfying (\ref{cond}) also satisfy
$r_m(\bb(z)) = r_m(\bb(-z))$.  According to~\cite{bdhm}, the
propagator for $\CO$ in the dual CFT is obtained by evaluating the
spacetime propagator between the corresponding points on the cutoff
boundary at $r_{\rm m}$ (also see~\cite{gkpw}).  In the limit
$\epsilon \rightarrow 0$, we expect the Green function computed this
way to scale to zero as the Weyl factor of the boundary metric
diverges.
 
Defining the cutoff boundary curve 
\begin{equation} 
\BB(z) \equiv  (\bb(z),  r_m(\bb(z)) ), 
\label{ccurve} 
\end{equation} 
the propagator for a scalar field of mass $m$ between points $\BB(\pm z)$  
should be given by 
\begin{equation} 
G(\BB(z),\BB(-z)) = \int \CD \CP ~ e^{i \Delta \, L(\CP)}. 
\label{pathint} 
\end{equation} 
(We will always be interested in large masses so that $\Delta \approx
m$.) Here we are summing over all particle paths between the two
boundary points and $L(\CP) = \int (-g_{\mu\nu} \dot{X}^\mu
\dot{X}^\nu)^{1/2}$ is the proper length of the path. (Defined this
way, $L(\CP)$ is imaginary for spacelike trajectories.)  As a check,
note that the action accumulated along a stationary trajectory should
be $E_0 t$ where $E_0$ is the lowest energy the particle can have.  A
scalar field of ``mass'' $m$ in $\ads{3}$ has lowest energy eigenvalue
$\Delta$~\cite{bkl} -- so the paths in (\ref{pathint}) are weighted by
$\exp(i \Delta \, L)$.  In the semiclassical WKB approximation, the
path integral localizes to its saddlepoints and is given by a sum over
geodesics
\begin{equation} 
G(\BB(z),\BB(-z)) = \sum_g e^{-\Delta \, L_g(\BB(z),-\BB(z))}. 
\label{sum1} 
\end{equation} 
Here $L_g$ is the (real) geodesic length between the boundary points. 
According to~\cite{bdhm}, in the large $N$ limit, (\ref{sum1}) is also 
the CFT propagator. 
 
To check this statement, we will reproduce the known Green function 
of $\CO$ (\ref{eqt1}) from (\ref{sum1}) in global $\ads{3}$. 
As discussed by Matschull~\cite{matschull}, equal-time geodesics of 
(\ref{adsmet}) are circle segments obeying the equation 
\begin{equation} 
\tanh\chi \, \cos(\phi - \alpha) = \cos(\beta), 
\label{geodeq} 
\end{equation} 
where the geodesic reaches the $\chi = \infty$ boundary at $\phi = 
\alpha \pm \beta$. Setting $\alpha=0$, and cutting off the spacetime 
at 
\begin{equation} 
r_{{\rm m}} = \tanh(\chi_{{\rm m}}/2) = 1 - \epsilon(t,\phi), 
\label{cutoff} 
\end{equation}  
the unique geodesic between the boundary points $(t,\pm 
\beta)$ intersects the cut off boundary at $\phi_m^\pm$ which 
are fixed by 
\begin{equation} 
\tanh\chi_{{\rm m}} \, \cos\phi_{{\rm m}}^{\pm} = \cos(\pm\beta). 
\label{phimax} 
\end{equation} 
Our convenient choice $\epsilon(t,\phi) = \epsilon(t,-\phi)$ implies 
that 
\begin{equation} 
-\phi_m^- = \phi_m^+ \equiv \phi_m. 
\end{equation} 
Then the length of a geodesic travelling between 
$(t,\chi_{{\rm m}},\pm\phi_{{\rm m}})$ is 
\begin{equation} 
L(\phi_{{\rm m}},-\phi_{{\rm m}}) = 2 \ln\left[ \sinh\chi_{{\rm m}} \, 
\sin\phi_{{\rm m}} + (\sinh^2\chi_{{\rm m}} \, \sin^2\phi_{{\rm m}} + 
1)^{1/2} \right]. \label{length1} 
\end{equation} 
So, to leading order in $\epsilon(z) \equiv \epsilon(\bb(z))$, the 
geodesic length between the points $\BB(\pm z)$ on the cut off 
boundary curve is 
\begin{equation} 
L(\BB(z),\BB(-z)) = 2 \ln\left(2\sin\phi(z) \over \epsilon(z)\right). 
\label{geodlen} 
\end{equation} 
Using (\ref{sum1}) and~\cite{bdhm}, the CFT propagator and 
the bulk propagator are equated to give 
\begin{equation} 
T(z) \equiv \ln G(\BB(z),\BB(-z))  
= -2\Delta \ln 
\left(2 \sin\phi(z) \over \epsilon(z) \right)  
\label{eqt2} 
\end{equation} 
in the $\epsilon \rightarrow 0$ limit, where the boundary metric is 
$ds^2 = (1/\epsilon(t,\phi)^2) ( -dt^2 + d\phi^2)$.  This is exactly 
right from the AdS/CFT perspective since (\ref{eqt1}) is defined on 
the Weyl rescaled cylinder $ds^2 = -dt^2 + d\phi^2$. 
 
\subsection{Stationary particles} 
\label{stationary} 
 
\begin{figure}[t]                                  
\begin{center}                                  
\leavevmode                                  
\epsfxsize=3in 
\epsfysize=1.5in                                  
\epsffile{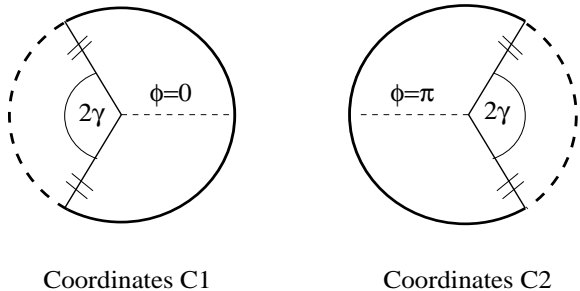}                                  
\end{center}                                  
\caption{Coordinate systems for a stationary point particle in 
$\ads{3}$.  The region $\pi + \gamma \leq \phi \leq \pi 
-\gamma$ and $-\gamma \leq \phi \gamma$ are removed in C1 and C2 
respectively.   The edges of the excised regions are identified. 
\label{statfig}} 
\end{figure}

Stationary point particles are introduced in three dimensional gravity 
by excising a wedge from the spacetime (\ref{adsmet}). (For 
example, see~\cite{mart,matschull} and references therein.)  A massive 
particle can only be stationary in (\ref{adsmet}) at $r=0$, and leads 
to the spacetime drawn in Fig.~\ref{statfig}.  The two coordinate 
systems displayed in the figure remove different wedges of the 
Poincar\'e disc: 
\begin{eqnarray} 
{\rm C1:}~~~~~   \pi + \gamma~>& \phi &> ~\pi - \gamma, \nonumber \\ 
{\rm C2:}~~~~~~~  -\gamma ~<& \phi &< ~\gamma. 
\label{statcoords} 
\end{eqnarray} 
In both cases the wedge boundaries are identified.  The dual CFT is 
still defined on the boundary at infinity, a cylinder whose spatial 
circle can always be chosen to have a period $2\pi$ by a coordinate 
transformation.  We would like to know how the presence of the 
particle is registered in the CFT. 
 
Since the asymptotic spacetime is locally AdS, local measurements on 
the spacetime boundary cannot tell us that there is a particle in the 
interior.  Nevertheless, the mass of the particle is available 
globally -- integrating the spacetime stress 
tensor of~\cite{adsstress} over the truncated range of $\phi$ gives 
\begin{equation} 
M = -{1 \over 8  G} + {\gamma \over 8 \pi G}. 
\label{mass1} 
\end{equation} 
The second term is the contribution of the particle. Equivalently, in 
the Chern-Simons formulation of three dimensional gravity, the mass of 
the particle is registered in the holonomy of the Chern-Simons gauge 
fields~\cite{mart}.  Within the AdS/CFT context, the computation that 
gives (\ref{mass1}) also gives the energy of the dual CFT state. 
Stationary particle spacetimes exist in the range $0 \leq \gamma \leq 
\pi$ with the associated masses $-1/8G \leq M < 0$.  The range $M 
\geq 0$ is occupied by the spectrum of BTZ black holes~\cite{BTZ}.

If multiple particles are placed in $\ads{3}$, a similar calculation 
will show that the expectation value of the dual stress tensor yields 
the total mass of the particles.  However, this is not enough data to 
locate the particles in spacetime or count them.  So we seek other 
ways of detecting point particles in $\ads{3}$.  We would like a 
quantity in the dual CFT that probes the interior of the bulk 
geometry.  The propagator is such an object, since it is determined by 
bulk geodesics running between boundary points. 
 
Following the pure AdS analysis, consider the coordinate system C1 and 
cut off the spacetime as in (\ref{regulate}).  As before, consider 
closed spacelike curves $\bb(z)$ satisfying (\ref{cond}) on the AdS 
boundary, and the associated cutoff boundary curve $\BB(z)$ 
(\ref{ccurve}).  Geodesics running between $\BB(\pm z)$ for small 
$\phi_1(z)$ will not pass through the wedge.  Their length is 
therefore given by the $\ads{3}$ result (\ref{geodlen}): 
\begin{equation} 
L(\BB(z),\BB(-z)) = 2 \ln\left(2\sin\phi_1(z) \over \epsilon(z)\right). 
\label{geodlen2} 
\end{equation} 
For large $\phi_1(z)$, the geodesics will run through the excised wedge. 
To analyze these trajectories it is convenient transform to 
coordinates C2: 
\begin{equation} 
r_2 = r_1 ~~~~~;~~~~~ t_2 = t_1 ~~~~~;~~~~~ \phi_2 = \phi_1 + \gamma 
~~ (\phi_1 > 0) ~~~~~;~~~~~ \phi_2 = \phi_1 - \gamma ~~ (\phi_1 <0 ). 
\label{transf1} 
\end{equation} 
The geodesics that intersect the identification in C1 are precisely 
those that miss it in C2 and vice versa.  The pure AdS result in C2 
gives the length of geodesics passing through the C1 wedge.  For some 
intermediate values of $\phi(z)$ there are two geodesics between the 
points $\BB(\pm z)$.  However, the sum (\ref{sum1}) is dominated by 
the geodesic with shortest, length which we will now focus on.   
 
By symmetry, the endpoints of the minimum length geodesics that do and
do not pass through the C1 wedge must be separated by the surface
\begin{equation} 
\pm \phi_1 = \bar{\phi} \equiv  {\pi-\gamma \over 2}. 
\end{equation} 
Let $\bar{z}$ be the parameter where $\phi_1(z) = \bar{\phi}$ .  The 
geodesics between $\BB(\pm z)$ for $0 \leq z \leq \bar{z}$ then 
foliate a spacelike surface on one side of the particle worldline, 
while $z \geq \bar{z}$ generates a spacelike surface on the other 
side.  It is easy to check that neither of these surfaces touches the 
particle worldline, but they meet at the AdS boundary at the angles 
$\phi_1 = \pm \bar\phi$.  Together they sweep out a spacelike surface 
that cuts through AdS, but has a tear in it which could be filled by a 
spacelike patch.  The worldline of the particle passes through this 
tear. 
 
In C1 coordinates we then have:  
\begin{eqnarray} 
0\leq \phi_1(z) \leq \bar{\phi}: ~~~~~~ L(\BB(z),\BB(-z)) &=& 2 
\ln\left(2\sin\phi_1(z) \over \epsilon(z)\right).  \\  
\bar{\phi} \leq 
\phi_1(z) \leq \pi - \gamma:~~~~~ L(\BB(z),\BB(-z)) &=& 2 
\ln\left(2\sin(\phi_1(z) + \gamma) \over \epsilon(z) \right). 
\end{eqnarray} 
Following the discussion for pure $\ads{3}$, 
\begin{equation} 
T(z) \equiv \ln G(\BB(z),\BB(-z)) = - \Delta \, 
L(\BB(z),\BB(-z)) . 
\end{equation} 
There is a kink in $T$ at the coordinate $z = \bar{z}$ at which 
$\phi_1(z) = \bar\phi$:  
\begin{equation} 
K = \left[{\partial T \over \partial z}\right]_{\bar{z}^+} 
- \left[{\partial T \over \partial z}\right]_{\bar{z}^-} 
= 2\Delta \left({\partial \phi_1(z) \over \partial z} 
\right)_{z=\bar{z}} \tan\left( {\gamma \over 2} \right)  
\label{kinks} 
\end{equation} 
So the mass of the particle in the bulk can be determined from the
strength of the kink in Green functions of generic scalar operators in
the dual CFT state:\footnote{The kink (\ref{kinks}) was computed in
the leading approximation. It is actually somewhat smoothed out by
interference between the two geodesics that exist between boundary
points in the vicinity of $\phi(z) = \bar{\phi}$, one of which enters
the identification.  Nevertheless, there is data in the propagator
that determines the mass of the bulk particle.  In the case of the
lightlike particles studied in the next sections there is a unique
geodesic between boundary points, leading to a genuine kink.}
\begin{equation} 
8\pi G M_{{\rm particle} } 
= \gamma = 2 \tan^{-1}\left( {K \over 4\Delta (\partial \phi_1(z) / 
\partial z)|_{z=\bar{z}} }  \right)  
\end{equation} 

\paragraph{Moral of the story:  } 
The family of geodesics between $\pm\BB(z)$ foliates a spacelike 
surface that cuts across the bulk spacetime.  This surface has a tear 
in it which surrounds the world-line of the stationary particle  
at the spacetime origin.  The kink in the CFT propagator 
(\ref{kinks}) arises because of the sudden jump between geodesics 
passing on one side and the other of the particle. In the next section 
we will use a similar strategy to locate moving particles in $\ads{3}$ 
by examining the resulting moving kinks in CFT propagators.

\section{Moving particles in $\ads{3}$} 
\label{moving} 
Boosting the stationary particle above leads to a moving particle 
spacetime.  Following Matschull~\cite{matschull}, we parametrize 
$\ads{3}$ as an SL(2,R) manifold via 
\begin{equation} 
{\bf x} = \cosh\chi \, (\cos t \, {\bf 1} + \sin t \, \gamma_0) + 
\sinh\chi \, (\cos\phi \, \gamma_1 + \sin\phi \, \gamma_2) 
\, , 
\end{equation} 
with gamma matrices 
\begin{equation} 
{\bf 1} = \pmatrix{1 & 0 \cr 0 & 1};~~~~ 
\gamma_0 = \pmatrix{0 & 1 \cr -1 & 0};~~~~ 
\gamma_1 = \pmatrix{0 & 1 \cr 1 & 0};~~~~ 
\gamma_2 = \pmatrix{1 & 0 \cr 0 & -1}. 
\end{equation} 
Here $(t,\phi,\chi)$ are the same coordinates appearing in
(\ref{adsmet}).  A boost along $\phi = 0$ is performed by the SL(2,R)
transformation
\begin{equation} 
{\bf x}' = u^{-1} \, {\bf x}\,  u;~~~~~  
u = \cosh(\xi/2)\, {\bf 1} - \sinh(\xi/2) \, \gamma_2. 
\label{boost} 
\end{equation} 
Applying this boost to the C1 and C2 coordinate systems for a 
stationary particle yields the corresponding descriptions of  a  
moving particle. 
 
The relation between the boosted and stationary (subscript $s$) 
coordinates is 
\begin{eqnarray} 
\cosh\chi \, \cos t &=& \cosh\chi_s \, \cos t_s, \nonumber \\ 
\cosh\chi \, \sin t &=& 
\cosh\chi_s \, \sin t_s \, \cosh\xi + \sinh\chi_s \, \cos\phi_s \, 
\sinh\xi,   
\nonumber \\ 
\sinh\chi \, \cos\phi &=&  
\sinh\chi_s \, \cos\phi_s \, \cosh\xi + \cosh\chi_s \, \sin t_s \, 
\sinh\xi,   
\nonumber \\ 
\sinh\chi \, \sin\phi &=& \sinh\chi_s \,\sin\phi_s. 
\label{boostcoords}  
\end{eqnarray} 
(The last equation is redundant, but is retained for convenience.) The 
particle that was stationary at the origin is now following the 
periodic trajectory 
\begin{equation} 
n \pi \leq t \leq (n+1) \pi:~~~~~~~~ 
\phi = n\pi,~~~ \tanh\chi = (-1)^n \, \sin t \, \tanh\xi 
\, , 
\end{equation} 
with $n$ any integer. The boost in (\ref{boost}) is an isometry and 
leaves the metric unchanged, but the identifications in 
(\ref{statcoords}) take a different form in the new coordinates.  In 
the new C1 and C2 coordinates, $t$ dependent wedges are removed behind 
and in front of the particle respectively, relative to its motion at 
$t = 0$. 
 
\begin{figure}[t]                                  
\begin{center}                                  
\leavevmode                                  
\epsfxsize=3in 
\epsfysize=1.5in                                 
\epsffile{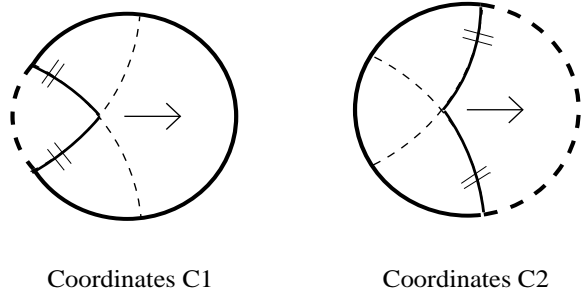}                                  
\end{center}                              
\caption{Coordinate systems for a moving point particle in 
$\ads{3}$. The edges of the excised regions are identified. 
\label{movfig}} 
\end{figure}                 
 
For purposes of illustration, we will examine the holographic
representation of lightlike particles moving through $\ads{3}$.  To
obtain such objects, take the limit $m\rightarrow 0, \, \xi
\rightarrow \infty$ while keeping $m\sinh\xi = \tan\delta$
fixed. Then, using (\ref{boostcoords}) for the boost (\ref{boost}) and
a similar relation for its inverse, the relation between the new C1
and C2 coordinates is:
\begin{eqnarray} 
\cosh\chi_1 \, \cos t_1 &=& \cosh\chi_2 \, \cos t_2, \nonumber \\ 
\cosh\chi_1 \, \sin t_1 &=&  
\cosh\chi_2\,\sin t_2 \left(1 +  {\tan^2\delta \over 2} \right) 
-{\sinh \chi_2 \, \cos\phi_2 \, \tan^2\delta \over 2} 
+ \sinh\chi_2 \, \sin\phi_2, \nonumber \\ 
\sinh\chi_1 \, \cos\phi_1 &=&  
\sinh\chi_2 \, \cos\phi_2 \left(1 - {\tan^2\delta \over 2} \right) 
+{\cosh\chi_2 \, \sin t_2 \, \tan^2\delta \over 2}  
+ \sinh\chi_2 \, \sin\phi_2, \nonumber \\ 
\sinh\chi_1 \, \sin\phi_1 &=& \sinh\chi_2 \, \sin\phi_2 
+ \tan\delta \, \cosh\chi_2( \sin t_2 - 
\tanh\chi_2 \, \cos\phi_2  ). 
\label{boostrel} 
\end{eqnarray} 
In this limit, the worldline of the particle is given by  
\begin{eqnarray} 
-\pi/2 \leq t \leq 0:~~~~~~~~ \phi=& -\pi,&~~~ \sinh\chi = -\tan t. 
\nonumber \\  
0 \leq t \leq \pi/2:~~~~~~~~ \phi=& 0,&~~~ \sinh\chi = \tan t. 
\label{nullparticle}  
\end{eqnarray}    
The boundaries of the excised wedge in C1 (C2) map to $\phi_2 = \pi$ 
($\phi_1 = 0$) in C2 (C1).  We will be interested in geodesics running 
between the endpoints $(t_e,\pm\phi_e,\chi_e)$ in either C1 or C2 
coordinates with $\sinh \chi_e \geq \tan t_e$ so that the endpoints 
are always further towards the AdS boundary than the bulk particle. 
In C1, there is a geodesic which does not pass through the 
identification if $\phi_e < \bar\phi$, where 
\begin{equation} 
\sin t_e - \tanh\chi_e \, \cos\bar{\phi}  = 0. 
\label{separ} 
\end{equation} 
Similarly, in C2, there is a geodesic which does not pass through the 
identification if $\phi_e > \bar\phi$ (in C1, these are the geodesics 
which do pass through the identification). The surfaces defined by 
(\ref{separ}) in C1 and C2 coordinates are identical; hence, there is 
a unique geodesic between any pair of endpoints.  
 
The coordinates (\ref{boostcoords}, \ref{boostrel}) have the virtue
that constant time slices are simply identifications of the Poincar\'e
disc.  However, constant $t_1$ and $t_2$ slices do not coincide.
Worse still, constant $t$ slices have a kink in them -- the vector
$\partial/\partial\phi$ is discontinuous at the identification.  We
avoid subtleties by working, as before, on some smooth spacelike slice
of the boundary; in C1 coordinates the curve is $\bb(z) =
(t_1(z),\phi_1(z))$ satisfying (\ref{cond}).  The smooth cutoff at
$r_m(t,\phi)$ in (\ref{regulate}) translates into
$\chi_{1m}(t_1,\phi_1)$ in C1.  So we have the cutoff boundary curve
$\BB(z) = (\bb(z),\chi_{1m}(z))$, following (\ref{ccurve}).  Similar
expressions in C2 are obtained by the coordinate transformation
(\ref{boostrel}).  As the curve $\BB(z)$ passes through the separation
surface defined by (\ref{separ}), geodesics between $\BB(\pm z)$ cross
over the bulk particle.
 
Geodesics between $\pm \BB(z)$ that avoid the C1 identification 
intersect the cut off boundary at $\pm\phi_{1m}$ where, following 
(\ref{phimax}), 
\begin{equation} 
\tanh\chi_{{\rm 1m}} \, \cos\phi_{{\rm 1m}} = \cos\beta. 
\label{phimax2} 
\end{equation} 
These geodesics have a path length (\ref{length1}) in C1 coordinates. 
The separation condition (\ref{separ}) determines some $\bar{z}$ such 
that, for $z>\bar{z}$, geodesics between $\BB(\pm z)$ intersect the C1 
identification.  By construction, these geodesics miss the C2 
identification and so have a length (\ref{length1}) in C2 coordinates. 
Setting 
\begin{equation} 
s_1(z) = \sinh\chi_{1m}(z) \, \sin\phi_{1m}(z);~~~~~ 
s_2(z) = \sinh\chi_{2m}(z) \, \sin\phi_{2m}(z), 
\label{defs} 
\end{equation} 
we have  
\begin{eqnarray} 
0\leq z \leq \bar{z}:& ~~~~    
L(z) = & -2 \ln\left[ s_1(z) + (s_1(z)^2 + 1)^{1/2} \right].  
\label{len1} \\ 
\bar{z}\leq z \leq \pi:& ~~~~ 
L(z) = & -2 \ln\left[ s_2(z) + (s_2(z)^2 + 1)^{1/2} \right].  
\label{len2} 
\end{eqnarray} 
%

As already established, $T(z) = - \Delta L(z)$ as $\chi_m \rightarrow 
\infty$.  The functions $s_1(z)$ and $s_2(z)$ are related by the last 
equation of (\ref{boostrel}), and, along with the separation condition 
(\ref{separ}), this shows that 
\begin{equation} 
s_1(\bar{z}) = s_2(\bar{z}) \equiv s(\bar{z}). 
\end{equation} 
So $T(z)$ is a continuous function of $z$. However, the derivative 
$dT/dz$ is discontinuous at $\bar{z}$.  Let 
\begin{equation} 
K = \left[ {dT \over dz}\right]_{\bar{z}^+} 
- \left[ {dT \over dz}\right]_{\bar{z}^-} 
= \left[ {dT \over ds} \right]_{s(\bar{z})} \,  
\left[ \left( {ds_2 \over dz} \right)_{\bar{z}^+} 
- \left( {ds_1 \over dz} \right)_{\bar{z}^-} 
\right]. 
\end{equation} 
Again using (\ref{boostrel}) and (\ref{separ}), $K$ vanishes only if 
\begin{equation} 
\left[ {d \over dz} ( \sin t_2(z) - \tanh\chi_2(z) \, \cos\phi_2(z) ) 
\right]_{z=\bar{z}} = 0. 
\end{equation} 
This fails since $z=\bar{z}$ is exactly the point where $\BB(z)$ 
intersects the vanishing locus of the quantity being differentiated. 
Thus, there is a kink in $T(z)$ at the surface (\ref{separ}) where the 
geodesics cross over the particle. 
 
\paragraph{Locating the particle: } 
In the holographic CFT, we can identify the position of the kink with 
the position of the moving particle in the bulk.  The arguments above 
showed that in C1 coordinates, there is a kink at 
\begin{equation} 
\cos \phi_1(\bar{z}) = {\sin t_1(\bar z) \over \tanh \chi_{1m}(\bar z)}  
\end{equation} 
As we remove the cutoff $(\chi_m \rightarrow \infty)$, the kink is at 
$\bar z$ where $\cos\phi_1(\bar z)= \sin t_1(\bar z)$.  So, recalling 
(\ref{nullparticle}), we can locate the particle in the bulk at the 
radial position 
\begin{eqnarray} 
\phi_1(\bar{z}) \leq \pi/2:&~~~~~~~~ \sinh \chi  &=  
-\tan\left(\phi_1(\bar{z}) - \pi/2)\right)  \nonumber \\ 
\phi_1(\bar{z}) \geq \pi/2:&~~~~~~~~ \sinh \chi  &= 
\tan\left(\phi_1(\bar{z}) - \pi/2)\right)   \, .  
\label{1position}  
\end{eqnarray} 
As in Sec.~\ref{single}, the geodesics running between $\BB(\pm z)$ 
sweep out a spacelike surface that cuts across the bulk spacetime. 
For a stationary particle, this surface had a spacelike hole in it 
through with the particle worldline passed.  Here, the surface 
foliated by the geodesics also has a tear in it that stretches between 
the two boundary points where the curve $\BB(z)$ intersects 
the separation surface (\ref{separ}).  However, it is readily shown 
that unlike the stationary case this tear can be patched by a null 
surface -- one edge of the tear is always ``later'' than the other 
along a null line.  The two sheets on either side of the tear are 
swept out by geodesics that do and do not pass through the C1 
identifications, and each sheet touches the null particle worldline at 
one point.  The formulae (\ref{1position}) locate the radial position 
of the earlier intersection. 
 
For simplicity, we imposed symmetry conditions like (\ref{cond}) and 
(\ref{regulate}) on our regulated boundary surfaces and on the curve 
$\bb(z)$.  In general, we simply have some closed, spacelike curve on 
the AdS boundary, which we parametrize in some arbitrary way as 
$\bb(z)$ with $-\pi \leq z \leq \pi$ .  To compute the propagator, we 
impose some cutoff $\chi_m(t,\phi)$, and study geodesics between the 
points $\BB(\pm z)$ ($\BB(z) \equiv [\bb(z),\chi_m(z)]$).  Each 
parametrization of the curve leads to a different family of geodesics 
sweeping out a different surface in spacetime.  The lightlike bulk 
particle will pass through a tear in this surface, touching the edges 
of the tear.  Kinks in the CFT propagators can be used to detect these 
locations. 
 
\paragraph{Relation to higher dimensions:} 
In \cite{garysunny}, the exact metric describing a lightlike particle 
in $\ads{d}$ for $d>3$ was constructed. This metric takes the form 
\begin{equation} \label{gspart} 
ds^2 = ds_0^2 + {p f(\rho) \delta(y_+) dy_+^2 \over (1+y_+ y_- - \rho^2)}, 
\end{equation} 
where $ds_0^2$ is the pure $\ads{d}$ metric. That is, the metric is 
pure $\ads{d}$ except on the surface $y_+ = 0$. If we write this 
metric in the usual global coordinates, as 
\begin{equation} \label{globaladsd} 
ds_0^2 = - \cosh^2 \chi dt^2 + d\chi^2 + \sinh^2 \chi (d\theta^2 + 
\sin^2 \theta d\Omega_{d-3}), 
\end{equation} 
then the surfaces of constant time are generated as surfaces of
revolution from a Poincar\'e disc; that is, the two-dimensional
subsurfaces given by choosing a pair of antipodal points on the
$S^{d-3}$ are Poincar\'e discs. We have singled out an axis in
(\ref{globaladsd}); assume this is the direction along which the
particle falls. Then in (\ref{gspart}),
\begin{equation} \label{yp} 
y_+ = {\cosh \chi \sin t - \sinh \chi \cos \theta \over 1 + \cosh \chi 
\cos t},  
\end{equation} 
and the Poincar\'e disc is changed by the addition of a delta-function 
at  
\begin{equation}  
\sin t - \tanh \chi \cos \theta = 0. 
\label{dcond} 
\end{equation} 
This in turn implies that the expectation value of the CFT stress 
tensor contains a delta function~\cite{garysunny}, the location of 
which, comparing (\ref{dcond}) and (\ref{separ}), is exactly analogous 
to the location of the kink in the CFT propagator we have found in the 
$\ads{3}$ case. 
 
\section{Colliding particles} 
\label{two} 
 
The techniques developed above for single particles can also be
applied to two colliding lightlike particles.  When the energy in the
collision is low, the result is the single particle spacetime of
Sec.~\ref{single}, but above a certain threshold a black hole is
created.  The representation of this process in the dual CFT is
central to understanding holographic descriptions of gravity.  The
only data available in the supergravity fields at infinity is the
total energy of the particles, and we seek ways of locating and
counting them.  Again, the propagator for an operator dual to a
generic $\ads{3}$ scalar field provides more information.  As above,
consider a one-parameter family of propagators whose associated
geodesics foliate a spacelike surface in the bulk.  Prior to the
collision we expect the null particle worldlines to pass through tears
in this surface.  Generically this will lead to two CFT propagator
kinks, arising as the bulk geodesics dominating the calculation cross
over the particle worldlines.  It is possible to have a single kink by
choosing a family of propagators whose associated geodesics hop over
both particles simultaneously.  But such special families have zero
measure in the space of possibilities we are considering, and so we
will ignore them.

\begin{figure}[t]                                  
\begin{center}                                  
\leavevmode                                  
\epsfxsize=1.5in 
\epsfysize=1.5in                                  
\epsffile{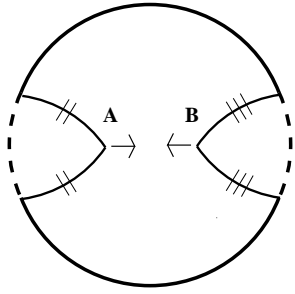}                                  
\end{center}                                  
\caption{Colliding particles in $\ads{3}$.  The edges of the excised 
regions are identified.}   
\label{2partfig} 
\end{figure}                                  
 
As discussed by Matschull~\cite{matschull}, a spacetime containing two 
particles is derived by excising a wedge for each particle from 
$\ads{3}$.  In the center of mass (COM) frame, particles approaching 
each other head-on can be displayed as in Fig.~\ref{2partfig}. 
Particles A and B enter the spacetime at $t=-\pi/2$ at $\phi = -\pi$ 
and $\phi = 0$ respectively and follow the trajectory $\sinh\chi = 
-\tan t$ (\ref{nullparticle}) until they collide at $t=0$.  In 
Fig.~\ref{2partfig}, each particle has a wedge excised ``behind" its 
trajectory -- if one particle was removed, the other would be in a C1 
coordinate system from the previous section.\footnote{For 
particle B, we have to reflect the C1 coordinates around the $\phi = 
\pi/2$ axis.  For example, angles in the range $0 \leq \phi \leq \pi$ 
are mapped to $\pi - \phi$.}  We are interested in geodesics running 
between the points $(t_e,\pm\phi_e,\chi_e)$ where $\chi_e \rightarrow 
\infty$.  For large (small) $\phi_e$, we expect the geodesics to enter 
the particle A (B) identification while for $\phi_e$ in the vicinity 
of $\pi/2$ both wedges should be avoided.  Following the single 
particle analysis and (\ref{separ}), we arrive at the two separation 
conditions 
\begin{equation} 
\sin t_e - \tanh\chi_e \, \cos \phi_A = 0;~~~~~~ 
\sin t_e - \tanh\chi_e \, \cos(\pi - \phi_B) = 0. 
\label{separ2} 
\end{equation} 
At any given $t_e$ and $\chi_e$, geodesics that start at $\phi > 
\phi_A$ ($\phi < \phi_B$) intersect the A (B) identification, while 
geodesics starting at $\phi_A < \phi < \phi_B$ pass between the two 
particles.  At $t=0$ the particles meet, and $\phi_A = \phi_B = 
\pi/2$. 
 
Again, consider closed spacelike boundary curves $\bb(z)$ satisfying 
(\ref{cond}) with $-\pi \leq z \leq \pi$.  Cut off the spacetime at 
$\chi_m(t,\phi)$ satisfying (\ref{regulate}) and define the associated 
curves $\BB(z)$ (\ref{ccurve}) on the cut off boundary. The separation 
conditions (\ref{separ2}) determine $z_A$ and $z_B$ at which geodesics 
between $\BB(\pm z)$ enter the A and B identifications.  Then 
arguments identical to Sec.~\ref{moving} show that there are kinks at 
$z_A$ and $z_B$.  At $t=0$ the two separation surfaces (\ref{separ2}) 
meet and so the two kinks join into a single one. 
 
\paragraph{Locating the particles prior to collision: } In the $\chi_m 
\rightarrow \infty$ limit, (\ref{separ2}) shows that the kinks occur
at $z_A$ and $z_B$ satisfying $\cos\phi(z_A) = \sin t(z_A)$ and
$\cos\phi(z_B) = \sin t(z_B)$.  We can thus locate the two bulk
particles at $\sinh\chi_A = \tan(\phi(z_A) - \pi/2)$ and $\sinh\chi_B
= -\tan(\phi(z_B) - \pi/2)$ prior to their collision.  The surface
swept out by the geodesics between $\BB(\pm z)$ has components
foliated respectively by geodesics passing behind A, between A and B,
and behind B.  Each component is spacelike and they meet at the
positions on the AdS boundary where the curve $\BB(z)$ intersects the
separation surfaces (\ref{separ2}).  However, in the bulk of
spacetime, the three components are separated by two tears through
which the particle worldlines pass.  As before, it can be shown that
these tears can be patched by a null surface -- the tear edges passing
between the particles are separated by null lines from the edges
passing behind the particles.  Each particle worldline touches the
edges of a tear, and we have used CFT kinks to locate the radial
position of the earlier intersection.

\paragraph{The flat space limit:} 
Flat space can be obtained from AdS by taking the limit $\ell 
\rightarrow \infty$ while keeping radial coordinate positions $r = 
\ell \tanh(\chi/2)$ fixed.  (See~\cite{flat} for discussions of the 
recovery of flat space from AdS.)  In this limit, the proper length 
between the A and B particles is simply $2 r_A$ and $\chi_A = \chi_B 
\rightarrow 0$.  So, the kinks will occur at $\phi(z) \approx \pi/2 
\pm \chi_A$. As $\ell \rightarrow \infty$ the proper length between 
any two boundary points is diverging.  However, since $z_A \rightarrow 
z_B$ in this limit, the ratio of the proper length between the 
locations of the kinks and the proper length of the boundary as whole 
tends to zero. This is symptomatic of a need for extreme precision in 
CFT measurements to resolve bulk objects at finite separation in the 
flat space limit.

\begin{figure}                                  
\begin{center}                                  
\leavevmode                                  
\epsfxsize=6.0in 
\epsffile{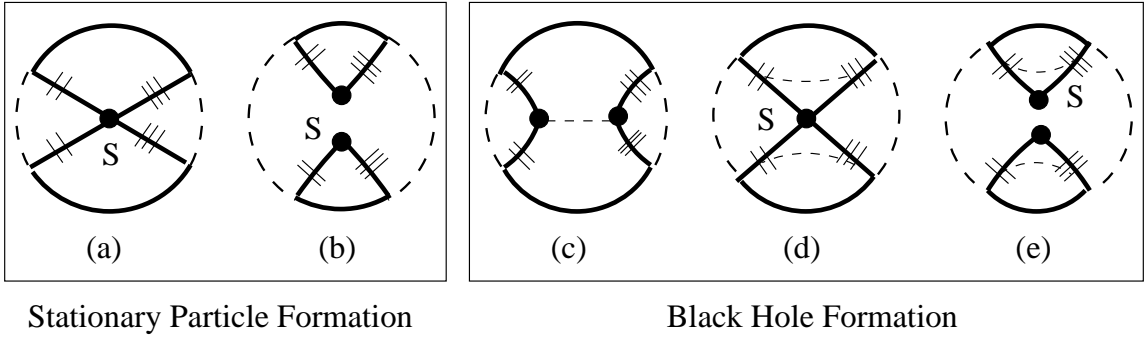}                                  
\end{center}                                  
\caption{(a) Particle collision below threshold,~~~(b) Evolution after 
collision below threshold,~~~   
(c) Horizon formation above threshold,~~~(d) Collision above 
threshold,~~~(e) Evolution above threshold}   
\label{collfig} 
\end{figure}                                  
 
\paragraph{Collision below the black hole threshold: }    
In (\ref{boostrel}), we derived the spacetime of a lightlike particle 
by boosting a mass $m$ particle with rapidity $\xi$, while keeping $m 
\sinh\xi = \tan\delta$ fixed.  The parameter $\delta $ is related to 
the energy of the particle, and it is readily shown that the head-on 
collision of two particles with $\delta \geq \pi/4$ creates a BTZ 
black hole~\cite{matschull}.  Below threshold, the collision produces 
the stationary particle spacetime in Sec.~\ref{single}, which we now 
analyze. 
 
Figs.~4a and 4b display the creation of a stationary particle.  The
particles A and B in Fig.~\ref{2partfig} enter the spacetime at
$t=-\pi/2$ and collide at the origin at $t=0$.  After this time the
spacetime in 4b is identical to that in Fig.~\ref{statfig}, but is
presented in skewed coordinates -- equal time slices in the two
figures do not coincide.  The two patches in 4b have identified edges
and the single stationary particle S at the tip of the patches follows
an oscillatory timelike trajectory that never reaches the spacetime
boundary, as appropriate to a massive particle~\cite{matschull}.
Prior to the collision, we have shown that there are two kinks in the
CFT propagator which approach each other.  The kinks arise because the
bulk surface foliated by geodesics between $\BB(\pm z)$ has two tears
in it surrounding the particle worldlines.  At $t=0$, the two tears
merge because there are no longer any geodesics passing between the
particles.  So, as manifest in Figs.~4a and 4b, there is only one kink
in the equal-time propagator when $t\geq0$, arising as the associated
geodesics switch between the two identifications.  The two-kink to
one-kink transition at $t=0$ marks the creation of the joint particle.

\subsection{Black hole formation} 
\label{bholesec} 
 
If the particles have a total mass greater than $1/8G$, they will 
form a black hole upon colliding.   We therefore have an exact metric 
describing the formation of a black hole. Three different 
coordinate systems are convenient for describing various aspects of 
the CFT representation of this process.

\begin{figure}                                  
\begin{center}                                  
\leavevmode                                  
\epsfxsize=2in 
\epsffile{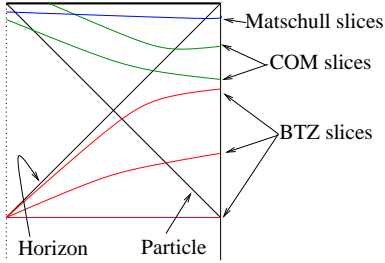}                                  
\end{center}                                  
\caption{COM, BTZ and Matschull's \btzm~slices of a black hole 
spacetime formed by collision of lightlike particles.    The slices 
are drawn at different times for pictorial clarity.   This is not 
really a Penrose diagram --   
the mappings between the coordinate systems do not preserve circular 
symmetry, and so we have simply displayed fixed angle slices in each.}  
\label{penrose} 
\end{figure}

In center of mass (COM) coordinates, the process is shown in Figs.~4c
-- 4f following~\cite{matschull}.  The particles enter the spacetime
at $t = -\pi/2$ and follow the trajectory in Fig.~\ref{2partfig}.  At
some later time $t = \tau - \pi/2$, an event horizon forms at the thin
dashed line in 4c.  It grows as the particles approach each other, and
the collision at $t=0$ in Fig.~4d creates a spacelike singularity
behind the horizon.  Successive COM spatial slices for $t > 0$
intersect this surface at the point S in Figs.~4d and 4e.  As time
passes, this point recedes towards the boundary of the Poincar\'e disc
along a spacelike curve.  It reaches the boundary at some $t=\tau$
which is the final spacetime point.  A picture of COM slices in the
global spacetime is shown in Fig.~\ref{penrose}.  It will transpire
that CFT propagators are easiest to compute in these coordinates.
However, the spatial slices have a kink in them at the
identifications, so they are not entirely natural from the CFT point
of view.

Smoother coordinates can be obtained by recalling that the absence of
gravitational degrees of freedom implies that even when the particles
are separate, the metric far from them can be written in BTZ
form~\cite{BTZ}:
\begin{equation} 
ds^2 = - N^2 \, dt_B^2 + r^2 \, d\phi_B^2 + {1 \over N^2} \, 
dr_B^2;~~~~ N^2 = r_B^2 - 8 GM. \label{btz} 
\end{equation} 
The holographic CFT is most naturally described in such coordinates,
in which the spacetime boundary is a smooth cylinder.  We will
construct the colliding particle spacetime in BTZ coordinates by
performing surgery on an eternal black hole.

\begin{figure}[t]                                  
\begin{center}                                  
\leavevmode                                  
\epsfxsize=4in 
\epsfysize=1.4in                                  
\epsffile{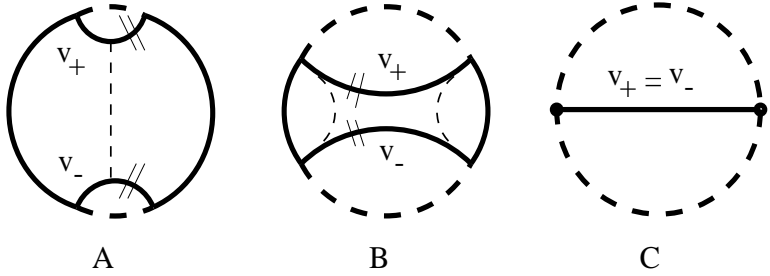}                                  
\end{center}                                  
\caption{\btzm : Poincar\'e disc coordinates for the BTZ black hole. 
(A) $t=-\pi/2$, (B) $-\pi/2 \leq t \leq 0$, (C) $t=0$ }  
\label{btz1} 
\end{figure}      
 
It is convenient to begin with an alternative form of the pure BTZ
metric, in which the spatial sections are identifications of the
Poincar\'e disc~\cite{matschull}.  Consider global AdS in the
coordinates (\ref{adsmet}) and identify the curves
\begin{equation} 
v^\pm:~~~~\tanh \chi \sin\phi = \mp \sin t \, \tanh\mu 
\label{adsident} 
\end{equation} 
for $-\pi \leq t \leq 0$.  The resulting spacetime is displayed in
Fig.~\ref{btz1}.  At $t=-\pi$ we have the collapsed geometry
associated with the past singularity.  As time passes, an
Einstein-Rosen bridge expands between two asymptotically AdS regions.
The horizons (thin dashed lines in Fig.~\ref{btz1}) come together,
meeting at $t=-\pi/2$ in analogy with equal time slices in four
dimensional Kruskal coordinates.  For $t>-\pi/2$ the Einstein-Rosen
bridge collapses again to a future singularity as shown in
Fig.~\ref{btz1}.\footnote{For $t<-\pi/2$ we have the time reverse of
Fig.~\ref{btz1}} We will call these coordinates \btzm,~while referring
to (\ref{btz}) simply as BTZ.
                  
\begin{figure}                                  
\begin{center}                                  
\leavevmode                                  
\epsfxsize=5.6in 
\epsfysize=1.5in                                  
\epsffile{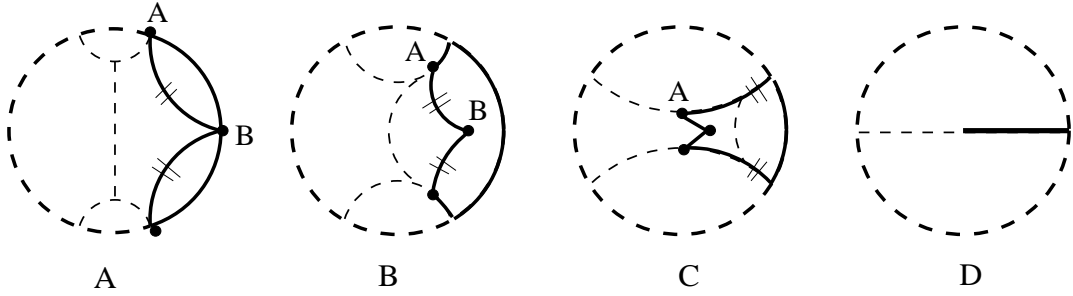}                                  
\end{center}                                  
\caption{\btzm~ coordinates for colliding particles.  (A) The
particles enter $\ads{3}$ at $t=-\pi/2$, (B) Both particles are
outside the horizon, (C) Both particles are inside the horizon, (D)
Collapse to a future singularity at late times}
\label{2partbtz} 
\end{figure}

We wish to drop two lightlike particles into $\ads{3}$ at $t=-\pi/2$. 
Matschull has shown how to present this spacetime in asymptotically 
\btzm~coordinates.  In Fig.~\ref{2partfig} each incoming particle was 
presented with a wedge removed ``behind'' its direction of motion 
(i.e., C1 coordinates from Fig.~\ref{movfig}). Now we simply excise a 
wedge from the \btzm~spacetime ``ahead'' of particle B in 
Fig.~~\ref{2partbtz}.\footnote{If we had excised this wedge from the 
full Poincar\'e disc we would have arrived at C2 coordinates in 
Fig.~\ref{movfig}.}  The resulting identification intersects $v^{\pm}$ 
at one point.  Particle A is placed at this point, as shown in 
Fig.~\ref{2partbtz}.  In these coordinates, particle B falls in along 
the trajectory $\phi =0, \tanh\chi = -\sin t$.    Near infinity,  the 
spatial slices of this collapsing spacetime are  identical to those of 
the  pure \btzm~black hole in Fig.~\ref{btz1}.

The coordinate transformation between \btzm~and BTZ coordinates is
obtained by relating both to a $(2,2)$ signature flat space in which
AdS is embedded as a hyperboloid.  This gives
\begin{eqnarray} 
{r_B \over r_+} \cosh(r_+ \phi_B) &=& -\cosh\chi \, \sin t, \nonumber \\  
{r_B \over r_+} \sinh(r_+ \phi_B)&=&  \sinh\chi \,\sin\phi, \nonumber \\  
\left( {r_B^2 \over r_+^2} - 1 \right)^{1/2} \cosh(r_+ (t_B + \pi/2)) 
&=& \sinh\chi \, \cos \phi,  \nonumber \\  
\left( {r_B^2 \over r_+^2} - 1\right)^{1/2} \sinh(r_+ (t_B+\pi/2))  
&=& \cosh\chi \, \cos t. 
\label{btzmtobtz} 
\end{eqnarray} 
Here $r_+ = \sqrt{8 \pi GM}$, and the points $(t,\pm\phi,\chi)$ in
\btzm~correspond to the points $(t_B,\pm\phi_B,r_B)$ in BTZ (with a shift
chosen so that $t=-\pi/2$ maps to $t_B=-\pi/2$).
 
The transformation between COM and BTZ coordinates can be found
similarly.  The equal-time slices of the three coordinate systems do
not coincide, and those of COM and \btzm~have kinks at the
identifications.  BTZ coordinates only cover the region outside the
horizon and are analogous to Schwarzschild coordinates for four
dimensional black holes.  Conveniently for us, the transformations
between the coordinates are symmetric under $\phi \rightarrow -\phi$
-- the points $(t,\pm \phi, \chi)$ in one of them is mapped into
$(t',\pm\phi',\chi')$ in another.

\subsubsection{Detecting particles inside an event horizon} 
 
We want to examine families of equal-time propagators as in previous
sections to study the CFT representation of black hole formation.  As
before, consider closed spacelike boundary curves $\bb(z)$ satisfying
(\ref{cond}) with $-\pi \leq z \leq \pi$.  Cut off the spacetime at
$\chi_m(t,\phi)$ satisfying (\ref{regulate}) and define the associated
curves $\BB(z)$ (\ref{ccurve}) on the cut-off boundary.  Then the
length of geodesics between $\BB(\pm z)$ as $\chi_m \rightarrow
\infty$ gives the CFT propagator between $\bb(\pm z)$.
 
It is easiest to begin in COM coordinates, which we have already
studied extensively.  Prior to the collision at $t=0$, we have shown
above that geodesics between $\BB(\pm z)$ sweep out a bulk surface
with two tears in it.  One component of this surface passes in between
the two particles.  As a result, we have shown that the associated CFT
Green function has 2 kinks until $t=0$ and that these kinks are
related to the positions of the colliding particles.  At later times,
the colliding particles have formed a spacelike singularity inside the
event horizon, and geodesics can no longer pass between them.  So, as
Figs.~4d and 4e make clear, there is only one kink in the propagator,
arising as the associated bulk geodesics switch between the
identifications, avoiding the spacetime singularity.  In the COM
slices with $t > \tau - \pi/2$, the particles are localized inside an
horizon; yet, we have just showed that their trajectories affect the
CFT propagator until they collide.  We have reached a remarkable, and
long sought-after~\cite{resol} conclusion: {\it there are simple
quantities in the holographic dual to $\ads{3}$ gravity that are
sensitive to the details of the matter distribution inside an event
horizon}!
 
The dual CFT is most naturally described on the smooth, cylindrical
boundary of conventional BTZ coordinates $\{t_B,\phi_B\}$.  So it is
interesting to locate kinks in the CFT Green function defined on
boundary curves $\bb(z)$ that coincide with synchronous BTZ
(\ref{btz}) slices, with the parameter $z$ equated to the BTZ angle
($z=\phi_B$).  It is helpful to first examine \btzm~geodesics running
between $(t_e,\pm \phi_e,\chi_e)$.  For small $\phi_e$ the geodesic
will pass ``behind'' particle B and miss the identifications in
Fig.~\ref{2partbtz}.  For large $\phi_e$ the geodesics will pass
``behind'' particle A and through the $v^\pm$ identifications.  In an
intermediate range of angles, the geodesics will pass between the
particles and through the wedge excised from \btzm~by the particles in
Fig.~\ref{2partbtz}.  Following our previous analyses, geodesics that
pass behind B are separated from those that pass between A and B by
the condition
\begin{equation} 
\sin t_{e}  - \tanh\chi_{e} \cos(\pi - \phi_e) = 0. 
\label{separ3} 
\end{equation} 
We will use (\ref{separ3}) to locate kinks in the CFT propagator 
defined on boundary curves $\bb(z)$ coinciding with synchronous BTZ 
slices. 
 
The qualitative form of the answer in the limit where the cutoff is
removed is clear from the COM discussion of colliding
particles. There, (\ref{separ2}) gave the location of the kinks, by
separating geodesics that do and do not pass through the
identifications in Fig.~\ref{2partfig} and Fig.~4c.  In the $\chi
\rightarrow \infty$ limit, this occurs at $\phi_{{\rm COM}} = \pm (t
-\pi/2)$ and $\phi_{{\rm COM}} = \pm (t + \pi/2)$, if we drop in
particles A and B from $\phi_{{\rm COM}} = -\pi,0$ at $t_{{\rm COM}} =
-\pi/2$.  That is, the associated kinks travel along lightcones. While
the relation between COM and BTZ coordinates is complicated, we expect
that in BTZ coordinates, the kinks at infinity should still travel
along lightcones.
 
To see this---and the form of the kink in the cut-off theory--- we use
the \btzm~to BTZ transformation (\ref{btzmtobtz}) and (\ref{separ3}) to
find the separation condition produced by particle B for BTZ geodesics
between $(t_{eB},\pm\phi_{eB},r_{eB})$:
\begin{equation} 
\left( {r_{eB}^{2} \over r_+^2} - 1 \right)^{1/2} \cosh(r_+
(t_{eB}+\pi/2)) = {r_{eB} \over r_+} \cosh(r_+ \phi_{eB}).
\label{separ4} 
\end{equation} 
By symmetry the condition for particle A is 
\begin{equation} 
\left( {r_{eB}^{2} \over r_+^2} - 1 \right)^{1/2} \cosh(r_+
(t_{eB}+\pi/2)) = {r_{eB}\over r_+} \cosh(r_+ (\phi_{eB} - \pi)).
\label{separ5} 
\end{equation} 
When we remove the cutoff by taking $r_e \rightarrow \infty$, the 
logic of previous sections shows that the CFT propagator on the BTZ 
cylinder has kinks at $\phi_B = \pm (t + \pi/2)$ and $\phi_B = \pm(t - 
\pi/2)$ as argued above.  The kinks move towards each other and meet 
when $t_B= 0 $ at $\phi_B = \pm \pi/2$.\footnote{In the regulated CFT 
defined on any cutoff surface at fixed $r_e$, the kinks still meet at 
$\phi = \pi/2$, but at a later time.}

As in previous sections, the family of geodesics is sweeping out some
bulk surface. At early times, the surface intersects each particle
worldline, producing two CFT kinks.  BTZ coordinates (\ref{btz}) are
analogous to 4d Schwarzschild coordinates, and only cover the region
outside the event horizon. Hence, in these coordinates, we never see
the particles cross the horizon and collide.  Nevertheless, geodesics
between some $(t_B,\pm \phi_B)$ must go out of the BTZ patch and
penetrate the horizon, since we know that geodesics between COM
boundary points $(t,\pm \phi)$ can penetrate the horizon and these are
mapped onto some $(t_B, \pm \phi_B)$.  The endpoints of the geodesics
which do this are to the past of the merging of CFT kinks in BTZ
coordinates. This is possible because the associated geodesics do not
remain at fixed time. They can bend into the future and some of them
penetrate the horizon, passing through the BTZ slices to the future in
the region the post-kink geodesics skip over. The two CFT kinks for
$t<0$ reflect where the surface foliated by the geodesics intersects
the particle worldlines, inside or outside the event horizon.  At late
BTZ times, the geodesics no longer intersect these worldlines.  The
kink in the CFT propagator then arises because of lensing by the BTZ
geometry -- the surface swept out by the associated geodesics skips
over the throat in the geometry, and the resulting tear is responsible
for the CFT kink.  If we consider a pair of particles that miss each
other and don't form a black hole, the kinks will approach each other
at $t=0$ and then separate again, returning to $\phi=0,\pi$ at
$t=\pi/2$.  This will happen because evolution in the gauge theory is
causal, so differences should only appear in regions that can receive
signals from both particles.
 
\section{Discussion} 
\label{conc} 
 
In this paper we have continued an ongoing discussion of the
holographic encoding of semiclassical gravitational backgrounds.  We
have learned some general lessons about the holographic AdS/CFT
correspondence.  The expectation values of CFT operators contain
enough information to completely characterize bulk solutions in a wide
variety of circumstances.  However, the relationships between CFT
expectation values and radial positions of bulk objects can be
complicated.  Furthermore, localized objects that are not built out of
mode solutions to the supergravity equations will be hard to
characterize using the expectation values of operators dual to
supergravity fields.
 
We argued that non-local CFT quantities, such as the propagator, can
resolve such localized objects and, as an example, studied collections
of point particles in $\ads{3}$.  We have shown that although the
asymptotic supergravity fields only contain data about the total mass
of the particles, the CFT Green function is able to enumerate and
locate them.  Related statements apply to spherical shells in
$\ads{5}$~\cite{dkk2, stevesimon}.  In both of these cases, the
localized object in the bulk has co-dimension one.  This is why the
bulk geodesics associated with the Green function are sufficient to
characterize it -- the geodesics probe the one extra dimension that is
is not occupied by the object, thereby locating it in that dimension.
Similarly, we might expect that objects of co-dimension $n$ can be
located in spacetime by transition amplitudes with $(n+1)$ particles.
 
One aim of this paper was to characterize the different states of a
holographic theory describing distinct bulk gravity solutions with
identical asymptotic fields.  We identified a CFT quantity which
depends on the bulk solution in the interior, but the story is far
from complete -- the ``propagator probe'' that we have used is not
fine-grained enough to resolve many questions of interest,
particularly in dimensions higher than 3.  We also expect that there
will not always be a unique gauge theory state associated with a
classical bulk solution (for example, there should be $e^{A/4}$ states
`associated' with a spacetime containing a black hole).  Our
techniques do not shed additional light on this issue.
 
In our approach, the process of black hole formation is visible in a
reduction in the number of kinks in CFT Green functions. Remarkably,
the CFT propagator is sensitive to the distribution of matter in the
{\it interior} of an event horizon.  In fact, our analysis of the
collision to form a black hole does not differ very much from
collision to form a stationary particle, until the singularity forms.

Note that the particles we used to form the black hole play a key
role in allowing us to explore its interior;  the geodesics that pass
between the particles are the ones that pass behind the horizon. If we consider
the pure BTZ spacetime, none of the spacelike geodesics between points
on one of the boundaries passes inside the black hole horizon. This
should not be a surprise; in the latter case, there is a second
asymptotic region, and the boundary conditions we obtain from the
gauge theory on one boundary are not sufficient to determine the spacetime inside the
black hole. 
 
This result is very suggestive -- from the CFT perspective, the view
of the black hole interior as a causally disconnected region with no
observable effect on the exterior is essentially misleading.  Explicit
study of holographic representations of black hole interiors is an
exciting direction for the future.

\vskip1in 
\centerline{\bf Acknowledgments} 
\medskip 
 
We have enjoyed discussions with Steve Giddings, Gary Horowitz, Nissan
Itzhaki, Per Kraus, Juan Maldacena, Nikita Nekrasov, Joe Polchinski,
Amanda Peet, Mark Spradlin, Andy Strominger and Lenny Susskind. The
work of {\small S.F.R.} was supported in part by NSF grant
PHY95-07065.  {\small V.B.}  was supported by the Society of Fellows
and the Milton Fund of Harvard University and by NSF grants
NSF-PHY-9802709 and NSF-PHY-9407194.  {\small V.B.} is grateful to the
ITP, Santa Barbara for its hospitality while this work was in
progress.
 

\end{document}